# WHEN THE LEFT HAND AND THE RIGHT HAND COOPERATE WITHOUT INFORMING THEIR OWNER

Brian W. Dodson
Research Foundations Division
Sandia National Laboratories[*]
Albuquerque, NM 87185-0836

**ABSTRACT**

A specific instantiation of classical correlation from entangled quantum resources can be established at a distance through the use of local measurements without classical communication. It is thereby possible to, e.g., develop a game in which distant entities can randomly combine forces on a course of action which is an element of a predetermined set of possible actions. When this protocol is used, none of the participants in the game knows which action has been chosen. The protocol is intrinsically secure in the same sense as a one-time pad. The entities must have a shared entanglement resource, be able to measure that resource in a predetermined reference frame, and know what they must do to carry out all of the possible actions.

An alternate title to this paper might be 'Correlated Choice of Classical Actions using only Quantum Entanglement and Classical Precommunication'. We use the term 'precommunication' to indicate that entities have shared classical information at some time before their choice is made. This might be done by an earlier physical meeting or through prior classical communication at a distance.

The idea investigated here is that correlated but mutually unknown actions of a set of agents which lead toward a goal chosen randomly from a set of potential goals can be established at a distance using only shared quantum entanglement, local measurements, and precommunicated classical information. The result is that a random selection of a set of classical actions can be made evident to all entities required to carry out the actions without classical communication. Any central agency will remain ignorant of the action chosen. This scheme is intrinsically secure in the same sense that a one-time pad is secure.

Imagine that Terorg, a terrorist organization, wants to execute a nefarious strike. Wanting to make defense against the strike as difficult as possible and having unimaginable resources, Terorg have prepared the ground for a set S of n possible strikes $S = \{s_1, s_2, \ldots, s_n\}$. However, Terorg have no preference as to which of the strikes is actually carried out – secrecy of the choice made and the identity of the agents involved is the primary consideration.

Alice and Bob are agents for the terrorist organization Terorg. (We will restrict our present discussion to the case of 2 field agents.) Neither Alice nor Bob know each other's identity or what the goals of the set of strikes are. They are merely to carry out a precommunicated sequence of actions on command which, when properly correlated, collectively serve to carry out a strike s which corresponds to one of the elements of S.

Alice and Bob have a controller Carl, who is their sole contact with Terorg. Carl's task is to insure that Alice and Bob both carry out a series of individual actions which insure that a randomly chosen member of the set of strikes is carried out. The maximum possible security is required, so as not to alert the targets of the strikes or to expose the agents. For simplicity assume that S has $2^n$ elements – therefore, each element of S can be described using n entangled two-state systems to establish a binary action integer which labels the individual actions leading to a particular strike.

The state of a finite-dimensional quantum mechanical system is a vector |R> in some complex n-dimensional Hilbert space $H^n$. (The quantum state is actually a ray, but this

distinction is unimportant here.) Let $H^n$ have an orthonormal basis $\beta = \{|\sigma_1\rangle, |\sigma_2\rangle, \ldots, |\sigma_n\rangle\}$. In that case, a state $|\Psi\rangle$ in $H^n$ can be expressed as $|\Psi\rangle = \alpha \Sigma \eta_i |\sigma_i\rangle$, where the $\eta_I$ are complex amplitudes and $\alpha$ is an overall normalization factor. We will be interested in states $|\Psi\rangle$ for which we can set all $\eta_I = \{0,1\}$ in the basis $\beta$. As usual, two quantum states are entangled when they are not separable.

Carl prepares a set of n entangled pairs of two-state systems. For simplicity we shall assume that n=1 for the current discussion. In Carl's computational basis $\Sigma$, the entangled pair has the state vector $|\Psi\rangle = 2^{-1/2} [\ |0\rangle_1 |0\rangle_2 + |1\rangle_1 |1\rangle_2\ ]$. He records this quantum state in two (decoherence-free) quantum memories, and gives one of the quantum memories to each of Alice and Bob. As they are maximally entangled as described above, it really doesn't matter who gets which particle. Alice and Bob also have equipment to align the orientation of their quantum memory basis to that of the basis $\Sigma$.

Carl now precommunicates Alice's actions. He informs Alice as to what actions she is to take when she measures the quantum state in the $\Sigma$ basis, obtaining a $|0\rangle$ state or a $|1\rangle$ state. Neither Carl nor Alice needs to know what two strikes are to be carried out by these two sets of actions. For maximum security, Dave, another member of Terorg, precommunicates to Bob the actions he is to carry out when he measures a $|0\rangle$ state or a $|1\rangle$ state. Introducing Dave yields the situation where neither Carl nor Dave individually knows all the actions required to carry out any of the strikes. This yields a significant increase in security, as knowing all the actions would probably allow them to accurately guess the identity of the possible strikes. In addition, no classical communication following the quantum state distribution is required with Terorg, so that protecting the identity of the agents does not depend on the security or anonymity of a classical channel. In this situation, Carl and Dave essentially carry out the function of chain cutouts, neither of whom has a complete picture of Terorg's intentions.

Terorg headquarters clearly know the set of potential strikes, but no one can tell them which strike will be carried out. Neither Carl nor Dave knows what binary number corresponds to which strike, or what binary number will result from the quantum state measurements carried out by Alice and Bob. Alice and Bob's measurements of the states stored in their quantum memories will determine which actions will be carried out, but not even Terorg headquarters can know which strike is to be carried out.

At some time before the earliest action of the set of strikes, Alice and Bob measure the state stored in their quantum memories, thereby obtaining their binary action number. There is a 50-50 chance of choosing either set of actions, and the quantum entanglement ensures that Alice and Bob choose correlated sets of actions by which they will combine forces on the same strike. The strike is chosen randomly in the measurement process, so that Alice and Bob combine forces on a strike the identity of which neither they nor Terorg headquarters can identify.

The procedure outlined above is reminiscent of the pseudo-telepathy of Brassard et al.[1], in which pseudo-telepathy is defined as an n-party game which cannot be won classically, but which can be won using quantum correlations at a distance.

Brassard et al. define a two-party game as a sextuple $G = <X, Y, A, B, P, W>$, where $X$, $Y$, $A$, and $B$ are sets, $P \subseteq X \times Y$ and $W \subseteq X \times Y \times A \times B$. $X$ and $Y$ are input sets, $A$ and $B$ are output sets, $P$ as a predicate on $X \times Y$ known as the promise, and $W$ as the winning condition, which is a relation between inputs $x \in X$ and $y \in Y$, and outputs $a \in A$ and $b \in B$ that has to be satisfied by $A$ and $B$ whenever the promise is fulfilled. $A$ and $B$ also win if $\{x, y\} \notin$ promise $P$.

At the outset of the game, $A$ and $B$ are allowed to discuss strategy and exchange any amount of classical information, including the value of random variables. If $A$ and $B$ are quantum players, they are also allowed to share unlimited amounts of entanglement. $A$ and $B$ are now separated so as to make further classical communication impossible.

$A$ and $B$ are each presented with questions $x \in X$ and $y \in Y$ respectively. $A$ and $B$ respond with answers $a \in A$ and $b \in B$. $A$ and $B$ win the round if $\{x, y, a, b\} \in W$, and $A$ and $B$ have a winning strategy if they continue to win as long as they have not exhausted the classical communication and quantum entanglement shared at the beginning of the game. A game exhibits pseudo-telepathy if there is no winning strategy when $A$ and $B$ are classical players, but does have a winning strategy provided $A$ and $B$ share a sufficient supply of quantum entanglement.

Although there is some apparent similarity between correlation of classical actions based on quantum entanglement and prior classical communication, these two processes are in fact rather different in operation and limitations. Perhaps the most obvious distinction is given by Brassard et al.[2], who prove that a minimal two-party pseudo-telepathy game requires an entangled quantum system of dimension at least 3x3. As the scheme described above requires a Hilbert space only of dimension 2x2, the underlying structure is clearly different from that of pseudo-telepathy.

It is clear that our protocol is intrinsically secure in the same sense as is a one-time pad. That is, there are no communications to intercept, so the only way in which in which Eve can attempt an attack is to physically copy or otherwise obtain the information in the pad. As the choice of strikes is not made until Alice or Bob measures the entangled state stored in their quantum memory, an attack requires that an additional state be entangled with Alice and/or Bob's entangled state in a manner so that Eve can obtain the information they have following their measurements. (We assume that if Eve can gain access to the entangled quantum states, she can also access the list coupling Alice and Bob's measurement results to the various strikes.)

Eve can try to use an entanglement-based attack to discover the binary action number shared by Alice and Bob by infiltrating Terorg headquarters and preparing sets of entangled triples rather than entangled pairs. Admittedly, this requires additional resources and equipment, but we will assume Eve is so clever that no one notices the extra equipment. Eve then keeps one of each entangled triple in her pocket for later reference.

Biseparable states[3] (e.g., $|101>+|110> = |1> \otimes [|01> + |10>]$) are of no use to Eve, as one qubit remains unentangled, so that one of the three participants would always measure a value uncorrelated to the binary action number. There are only two classes of genuine entangled triples, the GHZ state and the W state[4]. Ignoring normalization, the GHZ state is $|000> + |111>$ in the basis $\Sigma$, and the W state is $|011> + |101> + |110>$ in the same basis.

Preparation of W state triples is not useful for eavesdropping on the binary action number. Eve makes a measurement in basis $\Sigma$ of her particle from a W state triple. If Eve measures a $|0>$ state, she does know that Alice and Bob have both measured $|1>$ states. However, if her result is $|1>$, then the other particles are in a Bell state, and Eve gains no information as to Alice and Bob's measurement results. As a result, Eve cannot eavesdrop on Alice and Bob. At best, she can execute a denial of service attack, using the W-induced mixing of Alice and Bob's quantum states.

When a measurement in the basis $\Sigma$ is made of any particle in the GHZ triple, the result is a separable state ($|0>|0>$ if $|0>$ is measured or $|1>|1>$ if $|1>$ is measured). Accordingly, the post-measurement quantum state of the triple is no longer entangled. However, following a measurement in basis $\Sigma$ (assuming the particles remain effectively isolated from their

environment) the measured value and the remaining two particles remain completely correlated – they will all exhibit the same measurement in basis Σ. Since in the sample protocol described above Alice and Bob depend only on obtaining the same measurement result, this initial loss of entanglement does not invalidate the protocol. (Note that if the entanglement vanishes while the correlation in measurement results remains, entanglement monogamy[5] does not prevent this protocol from being used by three or more agents.) Thus, if Eve can produce GHZ triples and keep one member of each entangled triple in her pocket, she can eavesdrop on the binary action number (although one person's 'eavesdropping' might be another person's 'headquarters file copy'). In fact, Eve can identify the strike long before Alice and Bob carry it out, simply by measuring her particles first.

A physical attack on this protocol by an outside party (Wolf) can be made if Wolf knows the details of the protocol, the order of the particles relative to the desired binary action number, and the measurement basis Σ, and can obtain temporary custody (presumably illicitly) of one of Alice or Bob's set of particles. In this case Wolf can, in principle, use the N members of (say) Bob's set of particles as control states for a CNOT gate, and use additional particles in an appropriate quantum state as target states. E.g., if Alice and Bob's particles are in the same quantum state and Wolf uses |0> states as target states for the CNOT gate, the result is a set of GHZ triples of entangled particles. These can be used to reveal the key to Wolf as described above. This is analogous to stealing and copying a one-time pad.

The quantum correlation protocol described above does, in a perfect world, allow Alice and Bob to share a binary action number which is intrinsically secure in the usual meaning of that phrase. However, the world is not perfect – even when there is no quantum channel to exhibit noise, measurements can yield incorrect results as the result of a wide range of errors. Accordingly, it is worth investigating if there is any means for Alice and Bob to reconcile their individual binary action numbers[6]. If we attempt to maintain the complete isolation of Alice and Bob, it appears that conventional reconciliation is not possible, as that would require existence of a quantum and/or a classical channel between them.

How long a string of bits can, with high probability, be sent using our protocol? This length is largely controlled by the single-bit error probability $\epsilon$ through Shannon's Coding Theorem[7]. This theorem provides a relationship between the number of bits n shared by Alice and Bob, the single-bit error probability $\epsilon$, and the number of bits r which must be communicated between Alice and Bob to correct their strings. If we require r<1, the result is a rough limit on the length n of Alice and Bob's error-free strings:

$$n < [-\epsilon \log_2 \epsilon - (1-\epsilon)\log_2(1-\epsilon)]^{-1}$$

To obtain long strings with large probability of agreement requires very small error rates. For example, if $\epsilon = 10^{-4}$, the probable length of error-free strings is <700. We do not expect such small error rates in quantum measurements, although the possibility does exist. Using a more realistic (at present) error rate of .01, the probable string length is less than 13. Even though it is likely that a binary action string will be short, it seems likely that single-bit error rates of less than about 0.001 will be required.

Another approach to the sharing of identical binary action numbers presumes that Alice and Bob can both apply algorithms to substrings of M bits and thereby distill, without communication, single bits a and b which have a higher probability of agreeing than do the initial substrings. This problem has been studied by Mossel and O'Donnell[8]. They prove that, for a binary symmetric noise model (essentially $P(a_i=b_i) = 1-\epsilon$, $P(a_i \neq b_i) = \epsilon$ with unbiased statistics), there is no procedure to perform such distillation. In fact, Yang[9] has shown that the maximum correlation between a and b is $1-2\epsilon$, which can in fact be achieved by choosing a single corresponding bit from each substring. The choice of a single bit with an error probability of $\epsilon$, of course, does not assist Alice and Bob to agree on a random sequence.

It is now clear that, unless short error-free strings or longer strings with some errors are acceptable for Terorg's purposes, it will prove necessary to have some classical communication after measuring the particles. In this case, the usual procedures for reconciliation can be used to correct errors between Alice and Bob's strings.

At first sight, the proposed protocol seems an unlikely candidate for real-world application. However, it does serve to point out, in a rather simple example, the power that quantum entanglement provides compared to purely classical processes. Additionally, the rapid technical progress toward, e.g., entangled clock synchronization and entangled satellite communication stations suggests that our current estimates on the practical limitations of quantum information technology are likely overly conservative.

# **References**

*Sandia is a multiprogram laboratory operated by Sandia Corporation, a Lockheed Martin Company, for the Department of Energy's National Nuclear Security Administration under contract DE-AC04-94-L85000

1. Brassard, G., Broadbent, A. and Tapp, A., "Quantum Pseudo-Telepathy", Foundations of Physics <u>35</u>, 1877 (2005).

2. Brassard, G., Methot, A.A., and Tapp, A., "Minimum Entangled State Dimension for Pseudo-Telepathy", Quantum Information and Computation <u>5</u>, 275 (2005).

3. W. Dür, J. I. Cirac, and R. Tarrach, "Separability and Distillability of Multiparticle Quantum States", Phys. Rev. Lett. 83, 3562 (1999).

4. W. Dür, G. Vidal, and J. I. Cirac, "Three Qubits can be Entangled in Two Inequivalent Ways", Phys. Rev. A, 62, 062314 (2000).

5. V. Coffman, J. Kundu, W.K. Wooters, "Distributed Entanglement", Phys. Rev. A<u>61</u>,052306 (2000).

6. G. Brassard and L. Salvail, "Secret-key Reconciliation by Public Discussion", EUROCRYPT '93, Lecture Notes in Computer Science, Vol. 765, pp. 410–423. Springer-Verlag (1994).

7. C. E. Shannon, "A Mathematical Theory of Communication", Bell System Tech. J. **27**:379, **27**:623 (1948).

8. E. Mossel and R O'Donnell, "Coin Flipping from a Cosmic Source: On Error Correction of Truly Random Bits", Random Structures & Algorithms 26(4), pp. 418-436 (2005).

9. Ke Yang, Ph.D. Thesis, Computer Science Department, Carnegie-Mellon University (2005).